# reconCTI: A Proactive Approach to Cyber-Threat Intelligence

**Mohammed Mahir Rahman**
Department of Computer Science and Digital Technologies
School of Architecture, Computing and Engineering, University of East London
London, United Kingdom
ORCID: 0009-0007-4167-1475

**Shahzad Memon**
Department of Computer Science and Digital Technologies
School of Architecture, Computing and Engineering, University of East London
London, United Kingdom
ORCID: 0000-0003-3354-5798

**Tauseef Ahmed**
Department of Computer Science and Digital Technologies
School of Architecture, Computing and Engineering, University of East London
London, United Kingdom
ORCID: 0000-0002-1850-3496

**Ameer Al-Nemrat**
Department of Computer Science and Digital Technologies
School of Architecture, Computing and Engineering, University of East London
London, United Kingdom
ORCID: 0000-0003-0725-3417

***Abstract*— The rapid advancement of information technology has introduced a noticeable shift from traditional offline practices to more efficient and interconnected online environments. This transition, while offering convenience, has also increased exposure to various cyber threats such as identity theft, impersonation, and phishing scams. Reconnaissance, or briefly known as information gathering, is a key stage for threat actors, often relying on open-source intelligence (OSINT) to collect sensitive and extensive data on targets. In response to this challenge, this study introduces *reconCTI*, a command-line tool built using Python for Linux systems. The tool is designed to search for sensitive data leaks across both surface web and dark web platforms. It allows users to input specific keywords, scan multiple sites at once, and then assess the findings by referencing the MITRE ATT&CK framework. The results are compiled into a threat report that also includes possible mitigation strategies. *reconCTI* is intended to support both cybersecurity professionals and individuals in identifying risks early and taking appropriate action.**

## I. Introduction

As highlighted by Lockheed Martin [1], reconnaissance or the act of gathering information, is one of the key initial steps in launching a cyberattack. In this stage, adversaries aim to collect detailed intelligence about their targets to uncover behavioural patterns and identify potential security gaps that could be exploited. This intelligence is often sourced from publicly available data, commonly referred to as Open-Source Intelligence (OSINT), which includes content from search engines, social media, online forums, public records, and data breaches. Because OSINT can be accessed legally and with minimal effort, it becomes a valuable resource for attackers to craft highly targeted and effective intrusions.

The same reconnaissance techniques can also serve a critical role in defence. Currently many cybersecurity responses remain reactive, typically responding only after a breach has occurred [2]. This reactive nature often leaves defenders with limited time to contain or recover from attacks, increasing the risk of significant impact. Alternatively, when used proactively, reconnaissance allows both individuals and organisations to detect early indicators of compromise. By identifying exposed data before it can be weaponised, defenders are better positioned to implement safeguards and reduce the likelihood of exploitation. In recent years, the darknet has emerged as a significant contributor to the threat landscape [3]. Accessible only through anonymizing networks such as Tor (the onion router), the darknet hosts underground forums, marketplaces, and communication channels where stolen data, hacking tools, and exploit kits are exchanged. It is also a primary destination for the distribution of leaked databases - often the result of breaches in organizations and platforms that fail to secure sensitive information. Databases found across both indexed and unindexed web sources often contain sensitive information such as personally identifiable data (PII), login details, internal communications, and proprietary documents. If exposed, these data can pose lasting risks to individuals and organisations, including identity theft, financial loss, and reputational damage. Analysing these sources is essential - not only to understand current security gaps but also to anticipate future threats by monitoring patterns in data breaches and threat actor activity. To help keep track of such cyber activities in a structured way, analysts commonly rely on the MITRE ATT&CK framework - short for Adversarial Tactics, Techniques, and Common Knowledge [4]. Developed by MITRE Corporation, this framework offers a globally acknowledged reference for categorising attacker behaviour across the various phases of a cyber intrusion. It allows defenders to understand how adversaries operate, which methods they use, and how these can be detected or blocked. Using this structured mapping, organisations can strengthen their defences, better assess threats, and prepare more efficient response strategies.

With the rising volume of exposed data available in both the surface and dark web, and the absence of integrated tools to monitor these environments, this project introduces *reconCTI* - a tool designed for proactive threat intelligence gathering. *reconCTI* enables users to search for specific keywords across various online platforms, retrieving potentially sensitive content linked to those terms. Once the data is collected, the tool carries out an automated threat assessment by aligning findings with the MITRE ATT&CK framework. This highlights relevant attacker tactics and techniques, and the system then compiles the analysis into a detailed report that includes potential risks and suggested actions. The main contributions and novelties of this paper are:

1. Development of a lightweight reconnaissance tool to assist in proactive cyber threat intelligence gathering and analysis.

2. Enabling simultaneous scraping of multiple web sources across the surface and dark web.
3. Implementation of threat analysis mechanisms mapping the collected data to the MITRE ATT&CK framework.
4. PDF-based automatic threat intelligence reporting system for each session that aids decision-making.
5. Dual execution mode (guided/commando) to ensure accessibility for both novice and advanced users.

By combining OSINT collection, darknet monitoring, and structured threat mapping, *reconCTI* primarily aims to enhance the proactive threat detection capabilities of its users.

## II. Related Work

Open-Source Intelligence (OSINT) has become an essential element in cyber threat detection, offering access to a vast array of publicly available information that can help identify and assess emerging security risks. As digital threats grow more sophisticated, OSINT enables both individuals and organisations to collect and analyse online data in a structured and scalable way [5]. However, the sheer volume of available information often presents difficulties in filtering out noise and extracting actionable insights [6]. Reconnaissance - typically the first step in a cyberattack, can be approached using passive or active methods. Passive reconnaissance gathers data without direct interaction with the target, while active reconnaissance involves engaging with systems to collect more detailed intelligence [7]. Though these techniques have traditionally been used by threat actors, cybersecurity professionals are increasingly adapting them for defensive use [6]. Detecting early signs of data exposure can allow for preventative measures that reduce the risk of exploitation.

Recent studies highly expressed the value of OSINT in early threat detection. Researchers like Tounsi and Rais [8], and Sabottke et al. [9], show that analysing open data sources can provide proactive insights. OSINT is widely adopted in the public sector, where it helps monitor digital patterns and anticipate threats [10]. Still, cybersecurity often remains reactive, with intelligence collected post-incident [2]. Google's report [11] noted 97 zero-day exploits were active before detection, stressing the need for proactive strategies. The dark web also plays a significant role in threat intelligence, despite its anonymous nature and data validation challenges [12]. Its encrypted peer-to-peer structure supports illicit activities, including data breaches and coordinated attacks [13]. Several reconnaissance tools have been developed to automate the OSINT process. SearchOL is a passive web scraper that retrieves data from multiple surface web search engines and offers additional features like IP geolocation and metadata extraction [14]. theHarvester is another widely used tool that collects emails and usernames from platforms such as Google and LinkedIn [15], while Shodan provides insights into internet-facing IoT devices through passive scanning techniques [16]. These tools are effective in gathering raw data but are limited in their ability to process this data into actionable threat intelligence. Furthermore, most of them are confined to the surface web and do not address the vast information stored on the dark web, where sensitive breaches and attacker activities are often discussed.

TABLE I. Comparison with other OSINT tools

| Capabilities | reconCTI | SearchOL | theHarvester | Shodan |
|---|---|---|---|---|
| Surface/Dark web search | Both | Surface only | Surface only | Surface only |
| Continuous Monitoring | N/a | N/a | N/a | Available |
| Threat intel/mapping | Available | N/a | N/a | N/a |
| PDF threat reports | Available | N/a | N/a | N/a |

Although automated scraping and passive intelligence-gathering techniques offer efficiency and scalability [17], there remains a significant gap in transforming collected data into structured threat assessments. The current tooling landscape lacks integration with threat-mapping frameworks like MITRE ATT&CK, and does not provide clear mitigation strategies or proactive alerts based on findings.

In summary, while existing OSINT tools are instrumental for reconnaissance, they largely serve as data collectors rather than complete threat intelligence systems. There is an evident need for integrated solutions that not only gather information from both surface and dark web environments but also analyse, correlate, and present that data in a format useful for proactive cyber defence - addressing both known and emerging threats.

## III. Methodology

This section outlines the architecture and implementation methodology of *reconCTI*. The tool supports both surface and dark web reconnaissance by scraping user-defined keywords from multiple websites simultaneously. Based on the retrieved data, *reconCTI* generates a threat analysis report by mapping findings to the MITRE ATT&CK framework and a local CVE database. The final output is compiled into a PDF report, helping users to proactively mitigate potential threats.

### A. Overview of Functionality

The core functionalities of *reconCTI* are:

- User Input Collection: A guided (prompted question) and commando (one-line command) mode for users to input keywords and configure scraping.
- Surface Web Scraping: Collects data from open-source websites using user-specified keywords.
- Dark Web Integration: Utilises the Tor network to scrape onion domains.
- Threat Analysis: Maps scraped results to the MITRE ATT&CK framework and a local CVE database.
- PDF Threat Report Generation: Summarises identified risks and suggests mitigation strategies.

### B. Functionality Breakdown and Implementation

1. User Input Module

The user interaction is handled via two modes:

- Guided Mode: Prompts the user through a step-by-step process.
- Commando Mode: Allows power users to enter flags in a single line.

This is implemented in `modes.py` using standard Python libraries like argparse, json and os. Input data is saved to `history.json` for later reference.

2. Web Scraping Engine

The surface web scraping is conducted using:

- requests for HTTP requests,
- BeautifulSoup from bs4 for HTML parsing,
- and concurrent.futures.ThreadPoolExecutor enables scraping from multiple websites concurrently

In `scraper.py`, it filters out unnecessary content and extracts matched keywords.

3. Dark Web Integration

If the user selects `dark web` search, *reconCTI* establishes a Tor connection using:

- stem for Tor controller operations,
- requests routed via socks5h://127.0.0.1:9050 for anonymity.

Implemented in `tor.py`, the Tor session is bootstrapped and used for onion domain access:

```
session.proxies = {
  'http': 'socks5h://127.0.0.1:9050',
  'https': 'socks5h://127.0.0.1:9050'
}
```

4. Threat Analysis Module

After scraping, the results are analysed in `threat_analysis.py`. This module:

- Loads cve.json and mitre.json from the dat directory,
- Matches keywords with CVE and MITRE mappings,
- Produces a JSON summary (temp_analysis.json) including threat categories and suggested mitigations.

5. PDF Report Generation

The final threat intelligence is formatted into a structured PDF in threat_report.py using:

- fpdf - Python library.

The report includes:

- Search configuration summary,
- Detected keywords,
- Associated CVEs and MITRE techniques,
- Suggested mitigations.

### C. Ethical Considerations

All reconnaissance activities performed using reconCTI strictly adhere to publicly available information and avoid intrusive or exploitative behaviour. Surface web scraping respects `robots.txt` restrictions and employs rate-limiting to prevent denial-of-service effects. For dark web exploration, only publicly indexed `.onion` directories and known marketplaces are accessed, without authentication or data insertion.

## IV. Experimental Setup

To evaluate the effectiveness and practical capabilities of the developed tool *reconCTI*, a controlled experimental environment was configured using Oracle VirtualBox. The experiments were conducted within a virtualised Kali Linux system hosted on a Windows 11 machine. The experimental phase followed a scenario-based methodology to test the tool's capabilities under different conditions. The websites used for the testing purpose were:

1. Query - Question and Answer Website: http://ruc4i7xn5qu5uc7fu2sc34r6xl55xhgvxbcs56t4ayvbqo2fmp4pehqd.onion/
2. GitHub Gist: https://gist.github.com/

In the first scenario, two pieces of user data - a name and an email address - were input simultaneously to perform a combined (and logic) search across darknet forums. This method facilitates a more concise output. The second scenario focused on surface web reconnaissance, using only an email address as input. This test demonstrated the tool's surface-level scraping efficiency and its ability to locate leaked credentials or personal data. The final scenario involved an independent dark web scan without specific input, where *reconCTI* was tasked with identifying leaked data such as credentials, internal documents, or personally identifiable information from underground forums.

## V. Results and Evaluation

This section will demonstrate the outputs of the tests, based on three different scenarios:

### A. Scenario 1

The first scenario involved searching two types of data using the AND mode. The AND operator ensures that the scraper only returns results where both data values are present within the same page. In contrast, the OR operator allows the tool to search for multiple data values independently across different pages.

TABLE II. Search Specifications for Scenario 1

| Data Type/s | Value/s | And/Or | Website/s |
|---|---|---|---|
| Name | Sheila Santiesteban | And | http://ruc4i7xn5qu5uc7fu2sc34r6xl55xhgvxbcs56t4ayvbqo2fmp4pehqd.onion/ |
| Email | sheila.emili@yahoo.com | | |

The initial interface of the tool is illustrated in *Fig 1* and the logo is created with ASCII art.

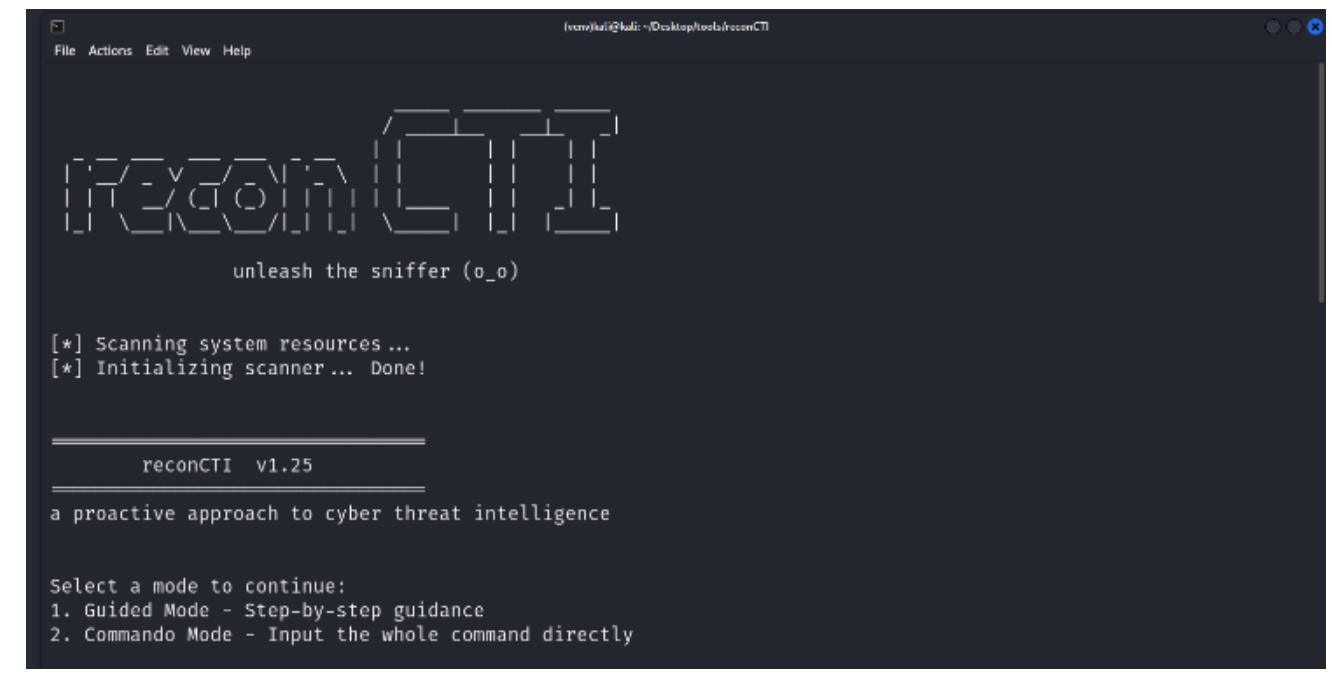

Fig. 1. Initial interface

The information were searched on a darknet forum website called "Query – Question and Answer Website". After entering relevant data, once the user states Tor links would be scraped, `tor.py` checks if the Tor connection is set-up. The input screenshots are provided in *Fig 2*. Before the scraping process begins, users are prompted to specify the desired recursion depth. By default, this is set to 2, meaning the scraper will crawl the root website link (e.g., http://test.com) and traverse no deeper than two additional subdirectories (e.g., http://test.com/1/2/). If the recursion depth is set to 4, the crawler will go as deep as http://test.com/1/2/3/4/ before halting scraping. Once the scraping and analysis are complete, the tool prompts the user to choose whether they would like to save the results in a separate file (shown in *Fig 3*). If confirmed, a file named `sc_result-n.json` is generated, where `n` represents the sequential number of the saved result file. This file will be used later for threat analysis.

```
File  Actions  Edit  View  Help

Enter your choice (1 or 2): 1

You have selected: Guided Mode

[Guided Mode] Please answer the following questions step by step:

Enter the type of data to search (e.g., Name, Phone, Passport): Name
Enter the Name: Sheila Santiesteban

Do you want to add more data? (yes or no): yes
Enter the type of data to search (e.g., Name, Phone, Passport): Email
Enter the Email: sheila.emili@yahoo.com

Do you want to add more data? (yes or no): no

Select search mode (AND for all keywords, OR for any keyword): And

Will you need to search onion links? (yes or no): yes
[Success] Tor is working! You are using the Tor network.

Add websites to search (comma-separated): http://ruc4i7xn5qu5uc7fu2sc34r6xl55xhgvxbcs56t4ayvbqo2fmp4pehqd.onio
n/

[Info] Proceeding to scrape the data ...
[?] Enter max depth for scanning (default 2):

[+] Starting the scraping process ...
```

Fig. 2. Input for Scenario 1

```
File  Actions  Edit  View  Help
it-onion-websites
[!] Failed to access http://ruc4i7xn5qu5uc7fu2sc34r6xl55xhgvxbcs56t4ayvbqo2fmp4pehqd.onion/?dwqa-question_cate
gory=shit-onion-websites: SOCKSHTTPConnectionPool(host='ruc4i7xn5qu5uc7fu2sc34r6xl55xhgvxbcs56t4ayvbqo2fmp4peh
qd.onion', port=80): Read timed out. (read timeout=10)
[*] Scraping: http://ruc4i7xn5qu5uc7fu2sc34r6xl55xhgvxbcs56t4ayvbqo2fmp4pehqd.onion
[!] Failed to access http://ruc4i7xn5qu5uc7fu2sc34r6xl55xhgvxbcs56t4ayvbqo2fmp4pehqd.onion: SOCKSHTTPConnectio
nPool(host='ruc4i7xn5qu5uc7fu2sc34r6xl55xhgvxbcs56t4ayvbqo2fmp4pehqd.onion', port=80): Read timed out. (read t
imeout=10)

[+] Results saved to scrape_results.json

[?] Do you want to save an additional results file? (yes/no): yes

[+] Results saved to sc_result-2.json

[✓] Scraping completed!

[?] Do you want to analyse the search results? (yes/no): yes

[✓] Using latest result file: sc_result-2.json

[+] Starting threat analysis ...

[✓] Threats detected. Report saved to temp_analysis.json

[+] Generating Threat Report PDF ...
[✓] Report generated: /home/kali/Desktop/tools/reconCTI/threat_report.pdf
[✓] Report opened in browser.
```

Fig. 3. Scraping session complete

Following this step, the user is asked if they wish to initiate an analysis of the scraped results. The latest `sc_result-n.json` file is then parsed by `threat_analysis.py`, which identifies and maps potential threats based on both local CVE mappings and MITRE ATT&CK framework references. If any threats are identified, a PDF report is automatically generated and displayed to the user, detailing the associated risks and recommended mitigations. The threat report (shown in *Fig 4*) identified two potential risks involving an email address and a name. The associated threats and corresponding mitigation strategies were provided within the report. Additionally, the report included references to the MITRE ATT&CK framework, detailing the relevant tactic, technique, and mitigation IDs.

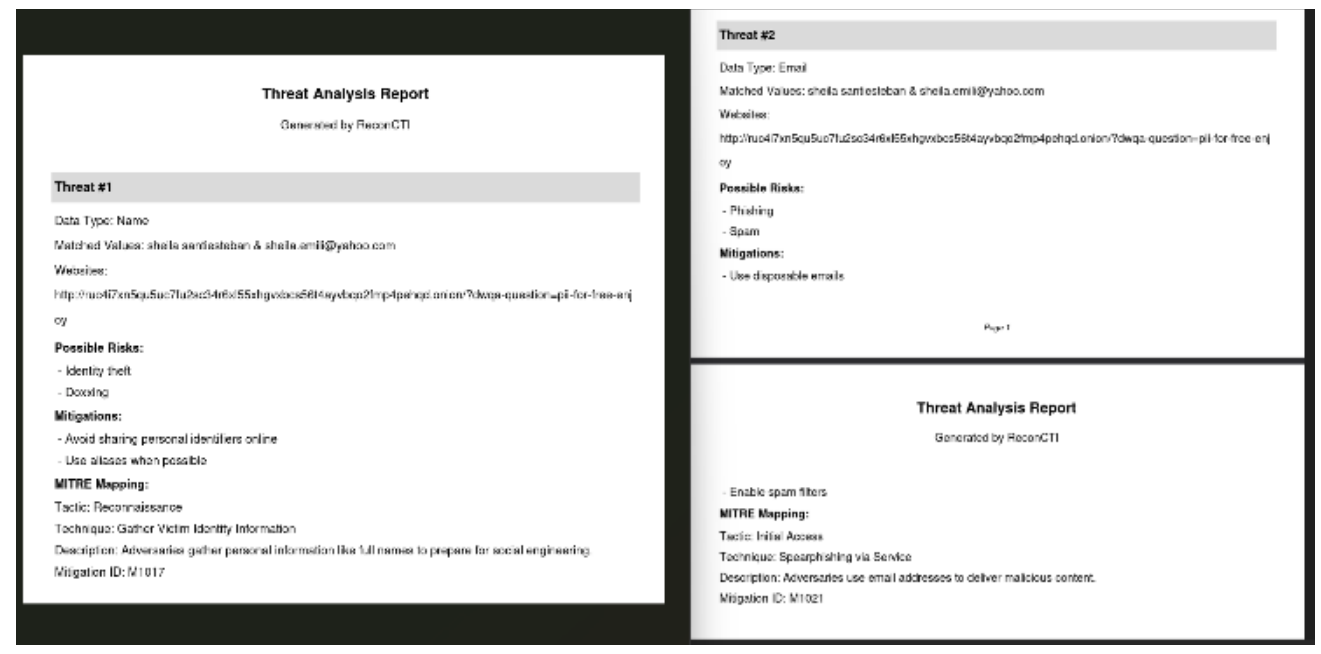

Fig. 4. Threat report pages 1 (left) and 2 (right)

This result demonstrates the capability of the *reconCTI* tool to successfully navigate through onion links and identify potential data leaks. The file `sc_result-2.json` illustrated in *Fig 5* contains further information about the scraping results and can be used to investigate the leaked data more thoroughly.

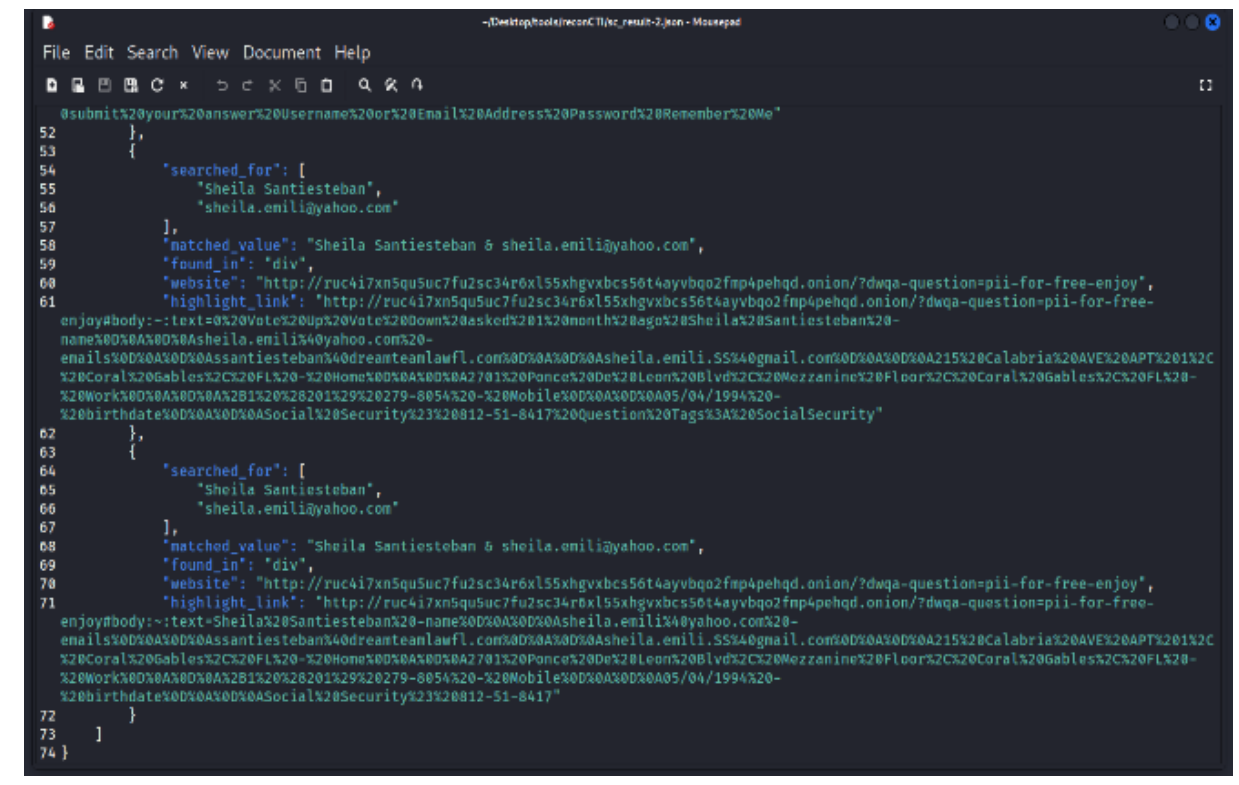

Fig. 5. Scrape result's file

Following the highlighted link in the document led to the original source of the information (shown in *Fig 6*). This source can potentially provide additional intelligence on how to address the identified threat.

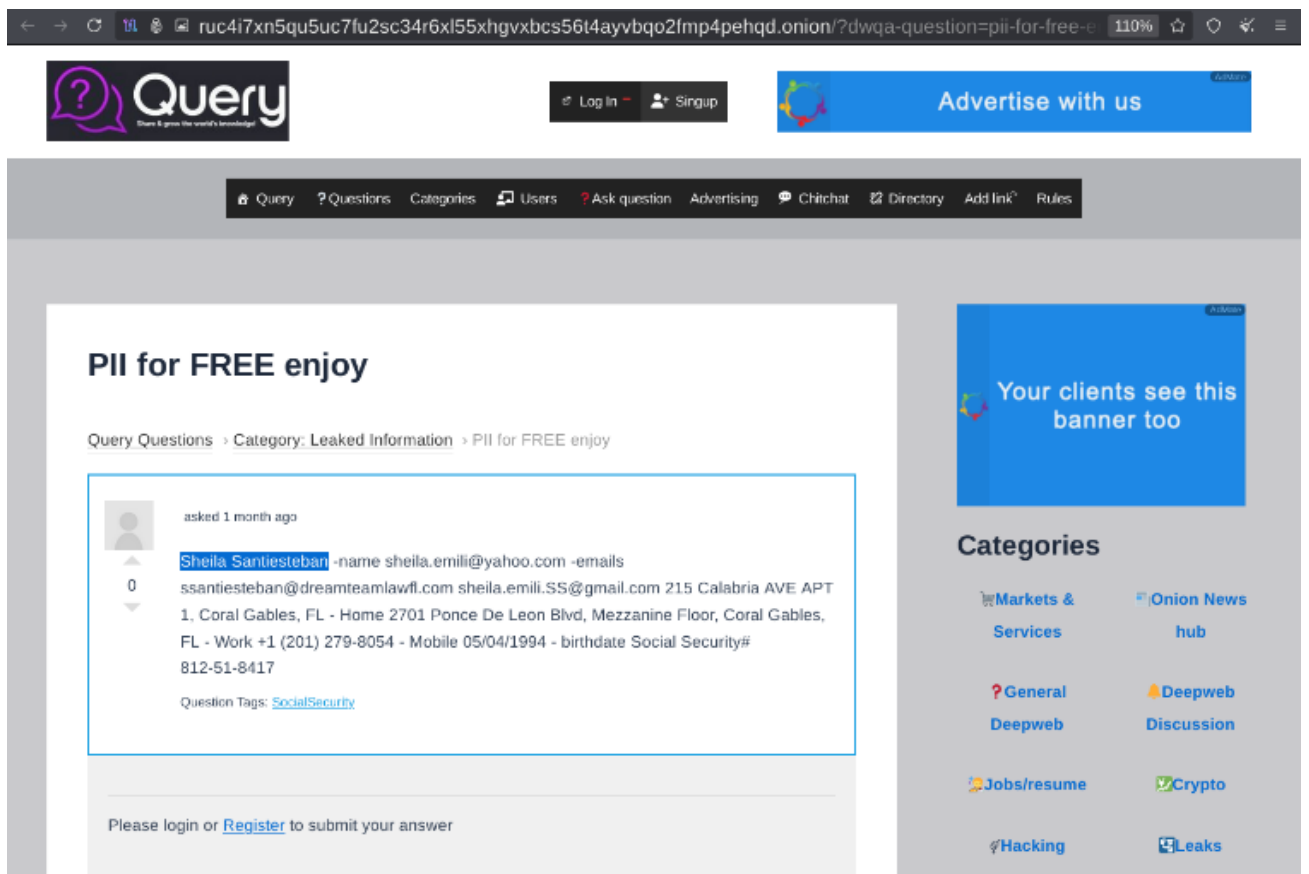

Fig. 6. Webpage where the data was found

## B. Scenario 2

The second scenario is based on surface web databases. A user's email address was scraped from a website using commando mode.

TABLE III. SEARCH SPECIFICATIONS FOR SCENARIO 2

| Data Type/s | Value/s | And/Or | Website/s |
|---|---|---|---|
| Email | arthurwelk83@whalebank.org | And/Or | https://gist.github.com/ |

Guided mode sequentially prompts the user for input and then carries out the scrape, whereas commando mode allows the user to input a single command to initiate the scrape. This method is particularly beneficial when searching for multiple values. A snippet of the input method is provided in *Fig 7*.

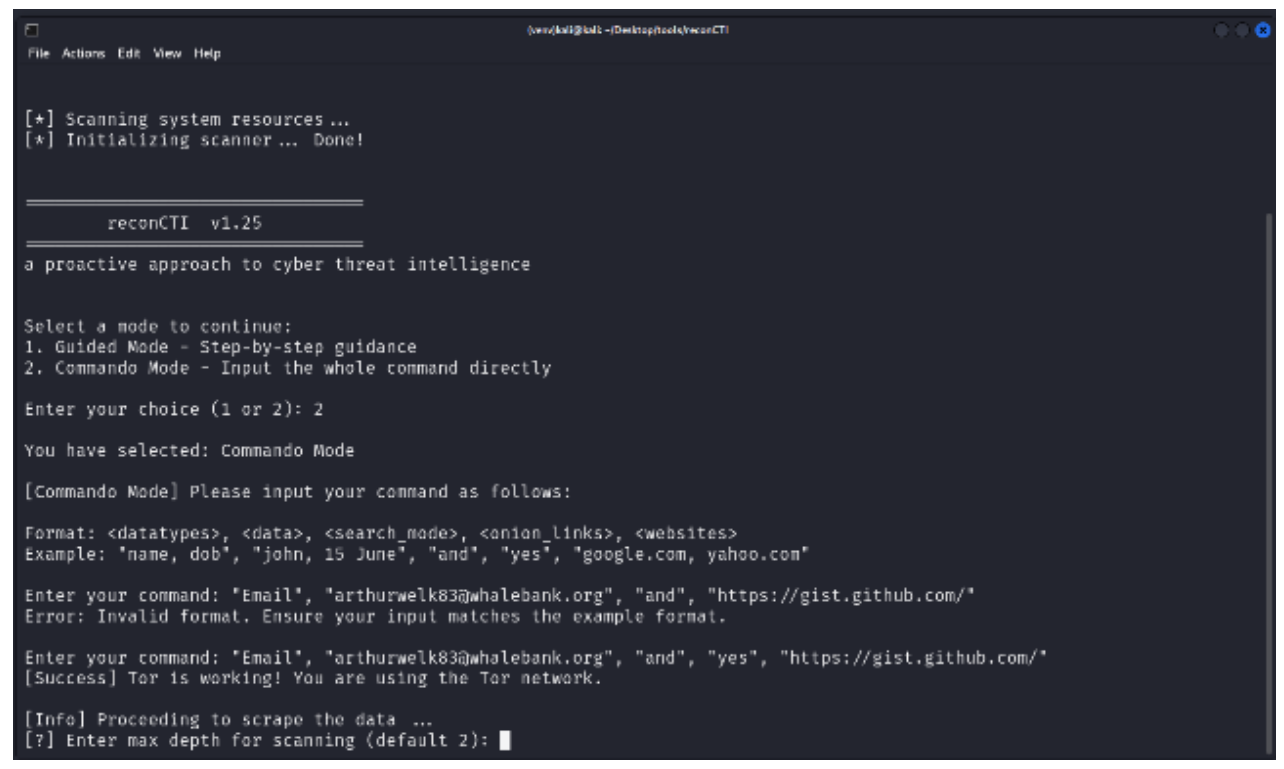

Fig. 7. Commando mode input method

The code is also designed to handle incorrect input, as shown in the snippet (*Fig 7*). Later, the threat report presented results including links where the data was found. This demonstrates that the scraper is capable of retrieving data from surface web links and providing valuable insight into the threats associated with the data leak.

**Threat Analysis Report**

Generated by ReconCTI

**Threat #1**

Data Type: Email

Matched Values: arthurwelk83@whalebank.org

Websites: https://gist.github.com/theodorehu95/2cf2e76d3bd6915e3902a5be264f244b#start-of-content, https://gist.github.com/theodorehu95/2cf2e76d3bd6915e3902a5be264f244b/revisions, https://gist.github.com/theodorehu95/2cf2e76d3bd6915e3902a5be264f244b#file-1000-leaked-email-addresses-txt, https://gist.github.com/theodorehu95/2cf2e76d3bd6915e3902a5be264f244b, https://gist.github.com/theodorehu95/2cf2e76d3bd6915e3902a5be264f244b/raw/2e684fb20986125ee192ae7d11edc1d611c56679/1000%2520Leaked%2520Email%2520Addresses.txt

**Possible Risks:**

- Phishing
- Spam

**Mitigations:**

- Use disposable emails
- Enable spam filters

**MITRE Mapping:**

Tactic: Initial Access

Technique: Spearphishing via Service

Description: Adversaries use email addresses to deliver malicious content.

Mitigation ID: M1021

Fig. 8. Threat report for Scenario 2

## C. *Scenario 3*

As part of this testing phase, a generic keyword was searched as shown in *Table IV*, with the aim of independently locating potentially valuable intelligence.

TABLE IV. SEARCH SPECIFICATIONS FOR SCENRIO 3

| Data Type/s | Value/s | And/Or | Website/s |
|---|---|---|---|
| Name/ Text | Onion Links that share data leaks for FREE | And/Or | http://ruc4i7xn5qu5uc7fu2sc34r6xl55xhgvxbcs56t4ayvbqo2fmp4pehqd.onion/ |

For the purpose of this test, the dark net forum Query was used again. The search input snippet is illustrated in *Fig 9*.

```
File Actions Edit View Help
a proactive approach to cyber threat intelligence


Select a mode to continue:
1. Guided Mode - Step-by-step guidance
2. Commando Mode - Input the whole command directly

Enter your choice (1 or 2): 1

You have selected: Guided Mode

[Guided Mode] Please answer the following questions step by step:

Enter the type of data to search (e.g., Name, Phone, Passport): Name
Enter the Name: Onion Links that share data leaks for FREE

Do you want to add more data? (yes or no): no

Select search mode (AND for all keywords, OR for any keyword): And

Will you need to search onion links? (yes or no): yes
[Success] Tor is working! You are using the Tor network.

Add websites to search (comma-separated): http://ruc4i7xn5qu5uc7fu2sc34r6xl55xhgvxbcs56t4ayvbqo2fmp4pehqd.onio
n/

[Info] Proceeding to scrape the data ...
[?] Enter max depth for scanning (default 2):
```

Fig. 9. Input snippet for scenario 3

After the scraping was completed, the result file was saved separately for further analysis. A snippet of the scraped results file is depicted in *Fig 10*.

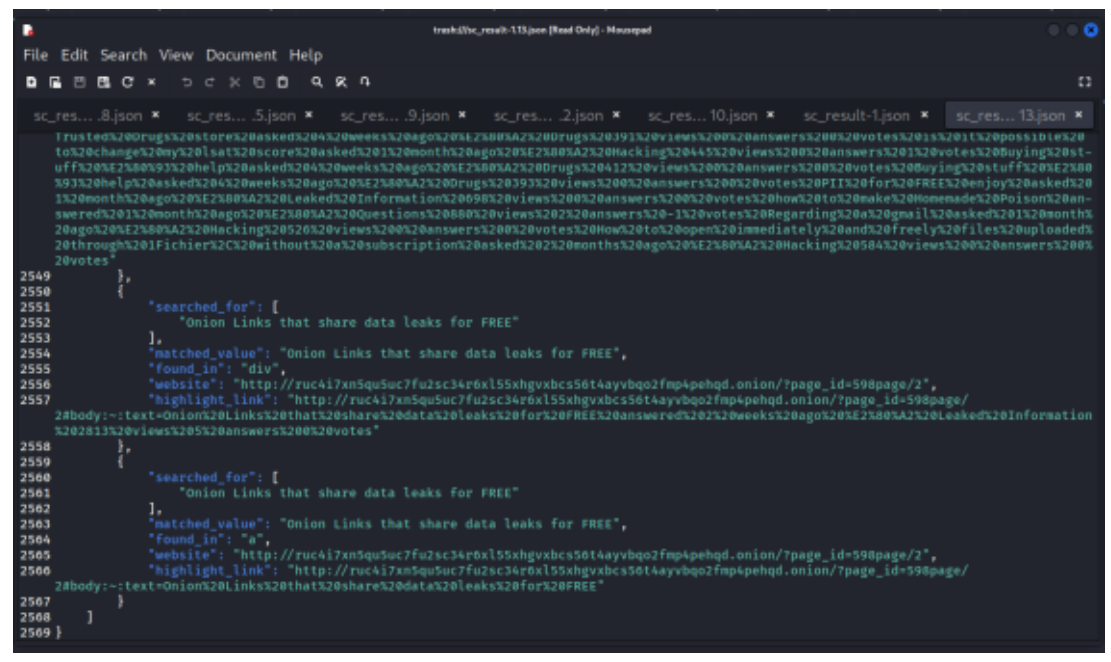

Fig. 10. Scrape result for scenario 3

There were multiple matches found throughout the search, and some of the most relevant matches provided highly valuable leads. OSINT or intelligence researchers often use similar methods to identify potential threat information. One of the matched links is shown below:

http://ruc4i7xn5qu5uc7fu2sc34r6xl55xhgvxbcs56t4ayvbqo2fmp4pehqd.onion/?dwqa-question=onion-links-that-share-data-leaks-for-free

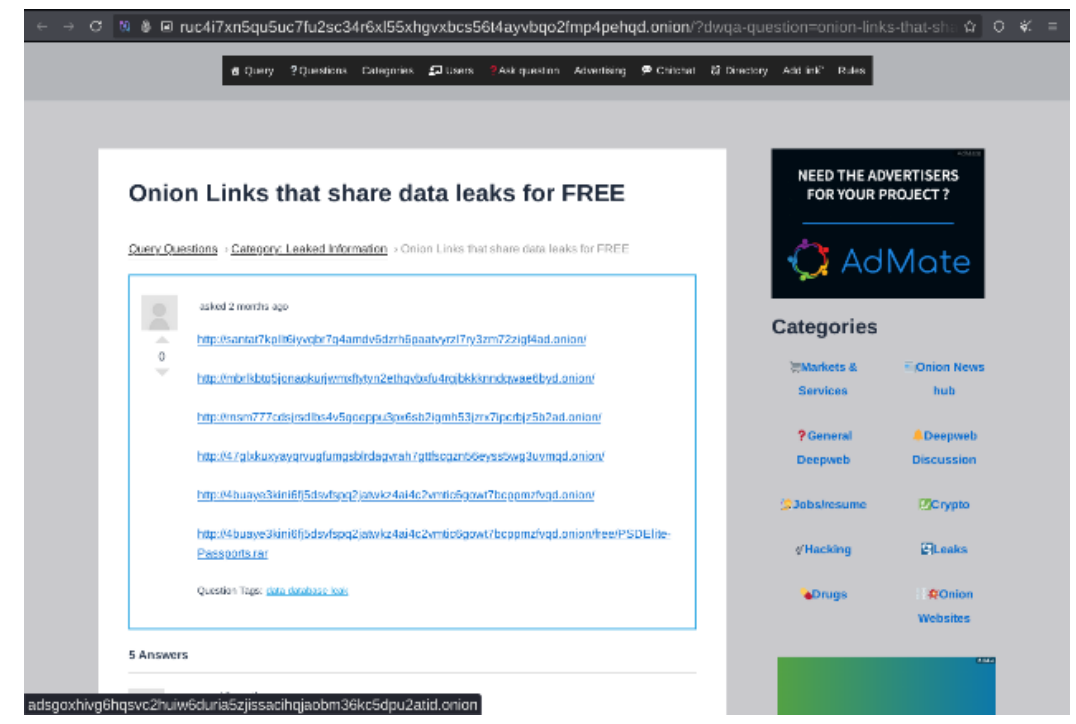

Fig. 11. The scraped result output

One of the forum discussions included several links that provided free access to onion pages containing leaked database information (shown in *Fig 11*). One of these links has been explored and is shown in *Fig 12*.

The link contained potential ransomware data belonging to a company and its users. Passwords, names, email addresses, and other sensitive information were also found to have been compromised. These were provided within the website by zip files and passwords to them (shown in *Fig 13*).

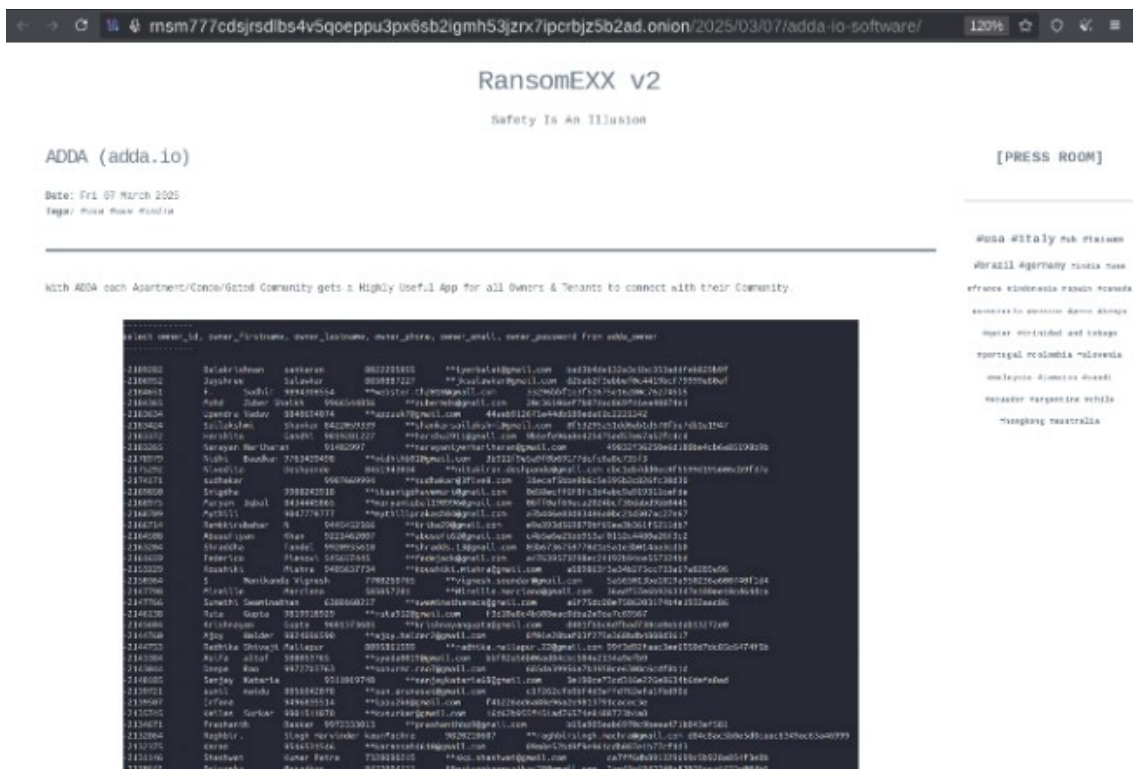

Fig. 12. Further investigation on the links found

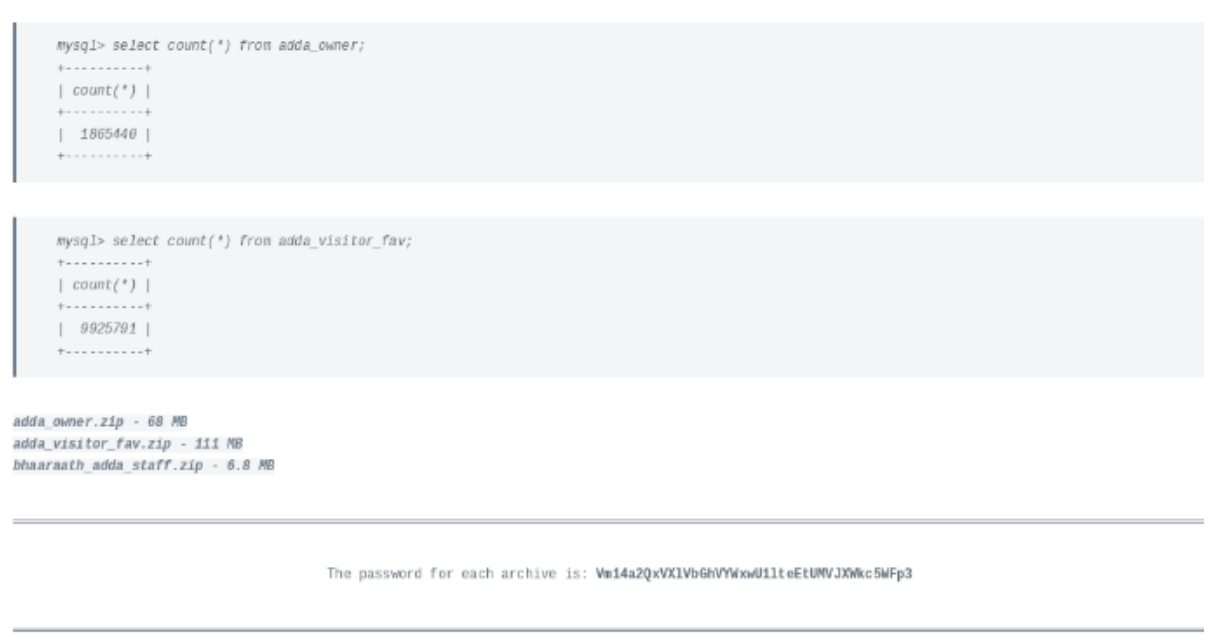

Fig. 13. Files with leaked data and password

This test demonstrates the strong capability of *reconCTI* to facilitate security research by automating the process of scanning for leaked information across darknet links.

### D. Performance Evaluation

In controlled tests using known leaks, *reconCTI* successfully identified all flagged data points, demonstrating strong detection capability. However, as the threat landscape evolves, new threats may go undetected due to limitations in the current database (further discussed in future work).

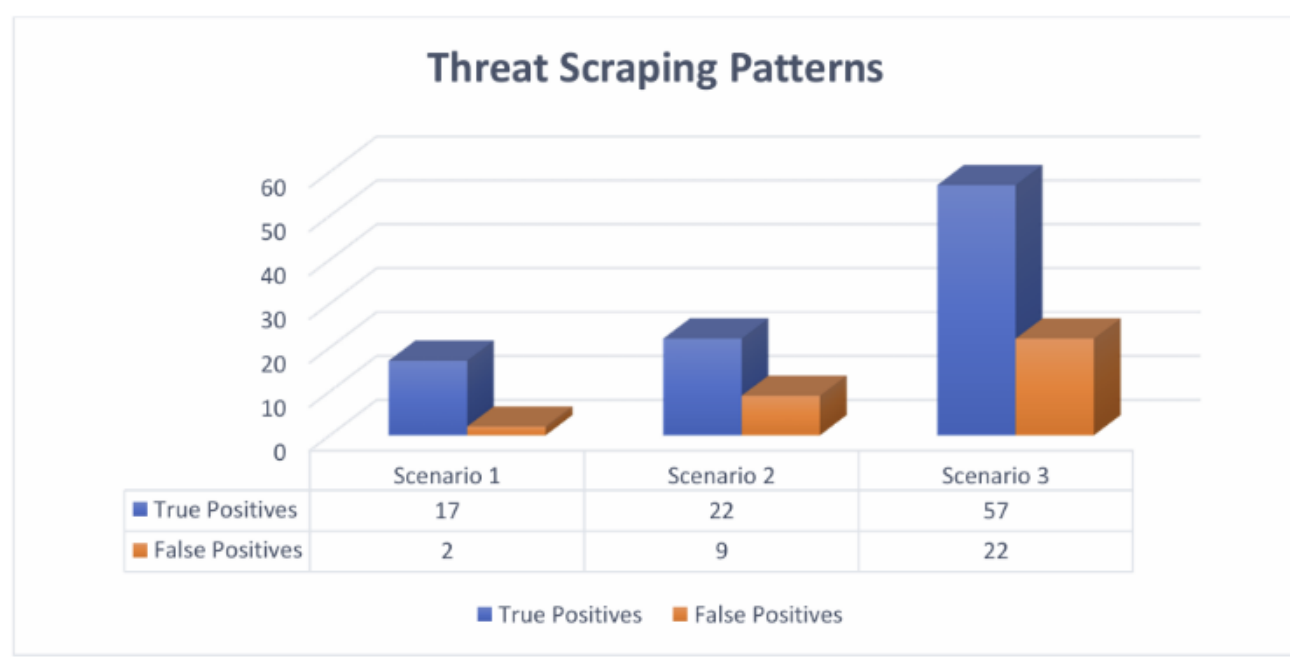

Fig. 14. Detection rates

Overall detection rates are shown in *Fig 14*. In Scenario 3, which involved generalized data, the false positive rate was moderately high - 22 out of 79 scraped links (*Fig 14*). Most false positives were from forum threads titled "Onion Links that share data leaks for FREE" with no actual content, or multiple pages of the same thread. To reduce such false positives, using the AND operator for more specific queries is recommended.

## VI. Conclusion and Future Work

This research project introduced *reconCTI*, a proactive reconnaissance tool capable of scraping both surface and dark web content to detect leaked data and generate threat intelligence mapped to MITRE ATT&CK. It achieved its objectives and showed promising results, with an average false positive rate below 30% in controlled tests. However, several limitations remain. The MITRE mapping relies on a static local database, and the scraper only handles static HTML content. Additionally, the keyword analysis uses exact-match logic without support for natural language processing or pattern recognition, limiting its ability to detect nuanced or indirect threats.

Future improvements include integrating live threat data via MITRE's TAXII 2.1, upgrading the scraping engine with dynamic content support (e.g., using Playwright or SeleniumBase), and enabling PDF/image parsing, CAPTCHA bypass, login automation and incorporating NLP for contextual threat analysis. In future research, the tool will be tested against a greater dataset and also involve qualitative interview approach with cybersecurity experts for feedback to improve the tool's intelligence depth and usability.

## Acknowledgment

I would like to express my sincere gratitude to my supervisors for their invaluable guidance, support, and feedback throughout the course of this project. I would also thank my parents for their continuous encouragement and support during this journey.